# Collective and synchronous dynamics of photonic spiking neurons

**Authors:** Takahiro Inagaki[1*], Kensuke Inaba[1*], Timothée Leleu[2], Toshimori Honjo[1], Takuya Ikuta[1], Koji Enbutsu[3], Takeshi Umeki[3], Ryoichi Kasahara[3], Kazuyuki Aihara[2, 4] & Hiroki Takesue[1]

* These authors contributed equally to this work.

**Affiliations:** [1]NTT Basic Research Laboratories, NTT Corporation, 3-1 Morinosato Wakamiya, Atsugi, Kanagawa, 243-0198, Japan

[2]Institute of Industrial Science, The University of Tokyo, 4-6-1, Komaba, Meguro-ku, Tokyo 153-8505, Japan

[3]NTT Device Technology Laboratories, NTT Corporation, 3-1 Morinosato Wakamiya, Atsugi, Kanagawa, 243-0198, Japan

[4]International Research Center for Neurointelligence, The University of Tokyo Institute for Advanced Study, The University of Tokyo, 7-3-1 Hongo, Bunkyo-ku, Tokyo 113-0033, Japan

*E-mail to: Takahiro Inagaki (takahiro.inagaki.vn@hco.ntt.co.jp) and Kensuke Inaba (kensuke.inaba.yg@hco.ntt.co.jp)

Nonlinear dynamics of spiking neural networks has recently attracted much interest as an approach to understand possible information processing in the brain and apply it to artificial intelligence. Since information can be processed by collective spiking dynamics of neurons, the fine control of spiking dynamics is desirable for neuromorphic devices. Here we show that photonic spiking neurons implemented with paired nonlinear optical oscillators can be controlled to generate two modes of bio-realistic spiking dynamics by changing the optical pump amplitude. When they are coupled in a network, we found that the interaction between the photonic neurons induces an effective change in the pump amplitude depending on the order parameter that characterizes synchronization. The experimental results show that the effective change causes spontaneous modification of the spiking modes and firing rates of clustered neurons, and such collective dynamics can be utilized to realize efficient heuristics for solving NP-hard combinatorial optimization problems.

## Introduction

Specialized hardware that performs brain-inspired information processing has achieved significant practical success in the fields of machine learning and artificial intelligence[1-4]. To provide more biologically realistic functions with artificial systems[5,6], various neuromorphic devices based on the spiking neural network (SNN) models have been developed[7-10]. Neurons communicate with nerve impulses, called spikes or action potentials, and the synchronization of the spikes can be useful for the signal processing performed in the brain[11-14]. The nonlinear properties of optical oscillators have been expected to be suitable for fast and energy-efficient implementations of the spiking neurons[15-22], but the photonic devices proposed so far have been generally limited in terms of diversity of their spiking dynamics. Since most nervous systems are constructed with various types of neurons, diversity and controllability of the spiking dynamics are important factors in building neuromorphic devices.

In the present study, it is demonstrated that a photonic artificial neuron can generate two different bio-realistic spiking modes that are changed spontaneously as a result of the synchronization within clusters of neurons. The artificial neuron was implemented with anti-symmetrically coupled degenerate optical parametric oscillators (DOPOs). The nonlinearity and phase bistability of the DOPOs were used to realize two spiking modes of class-I (saddle-node bifurcation) and class-II (Andronov-Hopf bifurcation) neurons that had been originally classified by A. L. Hodgkin[23] and characterized by different bifurcation structures[24]. These spiking dynamics can be controlled by tuning optical pump amplitudes of the DOPOs. We performed network experiments with 240 DOPO neurons and found that input signals from the correlating neurons can induce an effective change in the pump amplitude. The effective change occurs depending on the increase in the order parameter of synchronization, and it causes spontaneous changes in spiking modes and firing rates of the networked neurons. The experimental results showed that the self-tuning effect of collective spiking dynamics can be utilized for solving combinatorial optimization problems using methods related to self-organized criticality.

**Artificial spiking neuron with coupled DOPOs**

In this study, a photonic spiking neuronal network was developed by utilizing a network of DOPO pulses in a fiber ring cavity, which has been proposed for simulating an Ising spin network and solving combinatorial optimization problems[25-29]. The DOPO pulse is generated by a phase-sensitive amplifier (PSA) with a $\chi_2$ nonlinear material in the cavity[30-32]. Because degenerate parametric amplification is phase sensitive, the optical phase of each DOPO pulse takes only 0 or $\pi$ relative to the pump pulse above the threshold, and the optical amplitude of the bistable phase states can represent the positive and negative membrane potentials in the spiking neuron with sign-inversion symmetry. The spiking dynamics of each neuron is implemented by using a pair of coupled DOPO pulses (called $v$- and $w$-DOPOs) with respective coupling coefficients $J_{vw}$ and $J_{wv}$ as shown in Fig. 1a. The model of the *i*th neuron in a DOPO neural network is given as

$$\frac{dv_i}{dt} = (-1 + p_i)v_i - v_i^3 + J_{vw}w_i + \gamma \sum_j J_{ij}v_i + I_{\text{ext}} , \quad (1)$$

$$\frac{dw_i}{dt} = (-1 + p_i)w_i - w_i^3 + J_{wv}v_i + \gamma' \sum_j J_{ij}w_i , \quad (2)$$

where $v_i$ and $w_i$ represent in-phase components of DOPO amplitudes, and $p_i$ and $I_{\text{ext}}$ are its optical amplitude of pump and the external bias term, respectively. Quadrature components of DOPOs become negligible above the threshold due to the PSA[33-35]. The matrix $\{J_{ij}\}$ describes synaptic connections between the *i*th and *j*th DOPO neurons, and $\gamma$ ($\gamma'$) is a scaling factor of coupling strength for $v$- ($w$-) DOPOs (see supplementary S-A, S-B for more details).

A schematic diagram of the networked DOPO neurons is shown in Fig. 1b. We developed a network of 512 DOPO pulses based on time-domain multiplexing in a 1-km fiber cavity and opto-electronic feedback system[27]. A periodically poled lithium-niobate (PPLN) waveguide module and an optical band-pass filter were placed in the fiber cavity as a PSA. The continuous wave from a laser with a wavelength of 1536 nm was modulated by a lithium-niobate intensity modulator (IM1) into sequential pulses with 60-ps width and 1-GHz repetition frequency. The sequential pulses were amplified by erbium-doped fiber amplifiers and converted into 768-nm pump pulses by second harmonic generation (SHG) in the first PPLN waveguide (PPLN1). The pump pulses were converted into signal and idler waves through parametric down-conversion (PDC) in the second PPLN waveguide (PPLN2) in the 1-km fiber ring cavity. An optical band-pass filter with center wavelength of 1536 nm and passband width of 13 GHz was set behind the PPLN2 so that the transmitted light could satisfy the degenerate condition of the signal and idler waves. As a result of the interference of degenerate signal and idler waves, in which the 0 or π phase component relative to the pump phase was amplified most efficiently[36], phase-sensitive amplification could be obtained. When pumping the PSA was started, quadrature squeezed noise pulses were generated by spontaneous parametric down-conversion in the PPLN waveguide. The noise pulses were amplified in each cavity circulation by the PSA, which leads to the formation of time-domain multiplexed DOPO pulses. Since the pump pulse interval was 1 ns and the round-trip time of the 1-km fiber cavity was 5 μs, the cavity could accommodate more than 5000 DOPOs, 512 of which were used for this experiment.

Arbitrary all-to-all connectivity between those 512 DOPOs was implemented by using a measurement and feedback (MFB) scheme[37]. In each cavity circulation, a portion of each DOPO pulse was extracted with a 9:1 coupler, and the in-phase component was measured with a balanced homodyne detector (BHD). The local oscillator for the BHD was supplied from the continuous wave laser, which was used for preparing the pump pulse. The measured signals were then input into a field-programmable gate array (FPGA) module. The feedback signal for each DOPO pulse was calculated by using the input signals and a 512×512 coupling matrix with 8-bit connection weight resolution. The calculated feedback signals were imposed on the injection pulses by using a push-pull modulator and launched into the cavity at times synchronized with the target DOPO pulses. By repeating this procedure, it was possible to connect the 512 DOPO pulses in different time slots arbitrarily. In this study, thirty-two DOPOs were used as header pulses to monitor experimental conditions, and 480 DOPOs were assigned to simulate 240 DOPO neurons. The internal and external optical couplings of the 240 DOPO neurons could be controlled by changing the connection weights stored with the coupling matrix in the FPGA module.

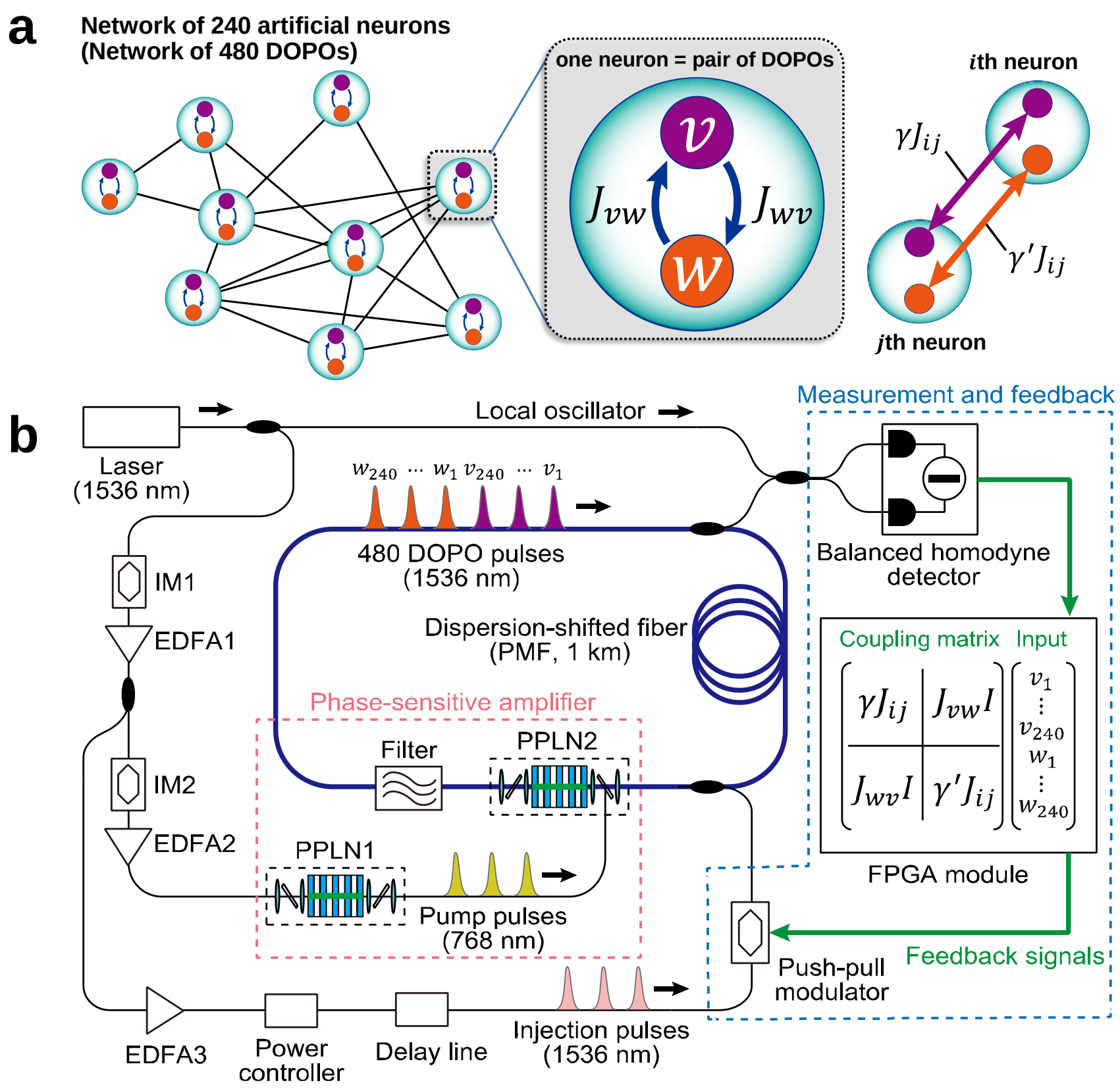

**Fig. 1** Experimental setup of a DOPO neural network. **a** A 240-node artificial neural network composed of 480 DOPOs with antisymmetric couplings ($J_{vw} = -J_{wv}$). **b** Schematic diagram based on time-domain-multiplexing in a 1-km fiber ring cavity. PPLN, periodically poled lithium-niobate; PMF, polarization-maintaining fiber; IM, intensity modulator; EDFA, erbium-doped-fiber amplifier; FPGA, field-programmable gate array

## Control of spiking dynamics

We realized spiking behavior of DOPO neurons that could be controlled by changing the pump amplitude. Time evolutions of coupled $v$- and $w$-DOPOs (blue and gray lines, respectively) were observed as shown in Fig. 2a. Constant pump amplitudes and an external bias linearly increasing from a negative value were applied to the DOPO neuron with antisymmetric coupling ($J_{vw} = -J_{wv}$). The following two parameters are defined: $P_i = (-1 + p_i)$ and $\omega_0 = \sqrt{-J_{vw}J_{wv}}$, where $\omega_0$ is natural spiking frequency at $P = 0$. Dimensionless notations are given as $\tilde{I}_{\text{ext}} \equiv I_{\text{ext}}/\omega_0^{3/2}$ for bias and $\tilde{X} \equiv X/\omega_0$ with $X = P, \gamma,$ and so on. For large $\tilde{P}$ (bottom panel of the figure), the interspike intervals gradually decrease from a very large value, meaning that firing rates gradually increase with increasing bias. For small $\tilde{P}$ (top panel), the interspike intervals and firing rates hardly change, while amplitudes gradually increase with increasing bias. These two kinds of behavior are very similar to those of the class-I and class-II neurons[23,24].

Spiking behavior of a single DOPO neuron was investigated by tuning pump amplitude $P$ and external bias $I_{\mathrm{ext}}$ with both $P$ and $I_{\mathrm{ext}}$ constant over time. Moreover, the model described by equations (1) and (2) was validated by comparing experimental results with numerical simulations (see Supplementary S-A). Change in spiking frequency $\tilde{\omega}$ as a function of $\tilde{P}$ and $\tilde{I}_{\mathrm{ext}}$ calculated by numerical simulations is plotted in Fig. 2b. As shown in Fig. 2c, the spiking frequencies of DOPO neurons with different operating parameters along three lines (A, B, and C) in Fig. 2b were experimentally measured. Our system could simulate up to 240 DOPO neurons simultaneously, and 80 neurons were assigned for obtaining data points corresponding to each line. Both the experimental measurements and simulations clearly show a sudden rise and fall of $\omega$ for line (A) and gradual increase and decrease of $\omega$ for line (B), characterizing the class-II and class-I neurons, respectively. These results show that the spiking mode can be switched between the classes II and I by tuning pump amplitude $P$. These behaviors are explained by the saddle-node bifurcation on a limit cycle (SNLC) (corresponding to class-I) and the Andronov-Hopf (AH) bifurcation (class-II) (see Supplementary S-B). The controllability of the neuron classes was observed by changing $P$ with $I_{\mathrm{ext}} = 0$ on line (C). When $P$ is changed from negative to positive, firing rate increases suddenly (AH) and then gradually decreases to zero (SNLC). Change in firing rate at $I_{\mathrm{ext}} = 0$ is approximately predicted as a function of $P$ as

$$\omega(P) = \omega_0\sqrt{1 - \frac{P^2}{8\omega_0^2}}, \tag{3}$$

where $0 \le \tilde{P} \le \sqrt{8}$ (see Method). The experimental results agree well with this function, and it was confirmed that the spiking mode of the DOPO neuron is seamlessly controllable between the class-I and class-II excitability.

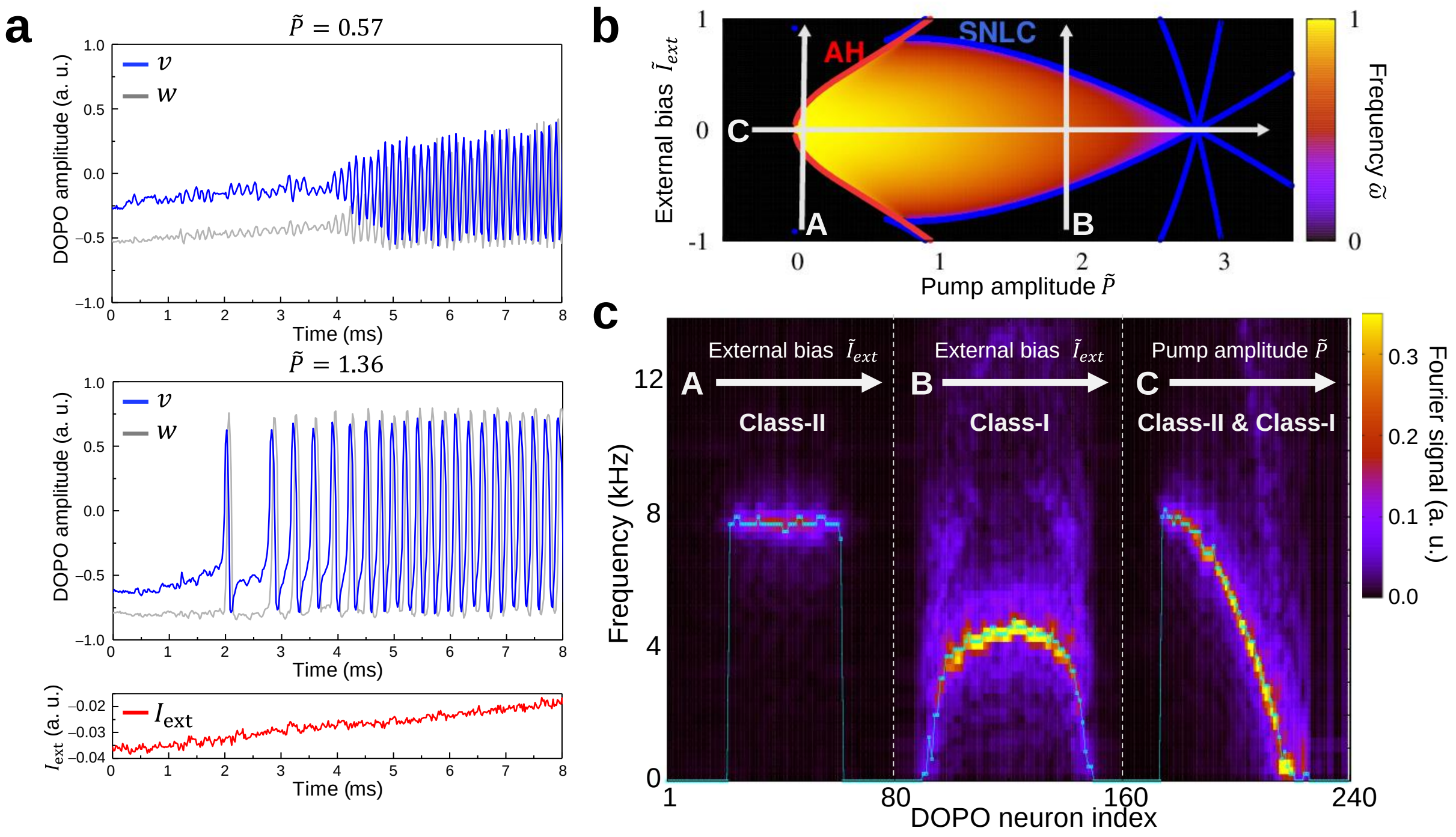

**Fig. 2** Class-I and II spiking modes of DOPO neuron. **a** Time evolutions of $v$- and $w$-DOPOs (blue and gray lines) with constant pump amplitudes ($\tilde{P}$ = 0.57 and 1.36) and an external bias increased linearly with time. **b** Phase diagram of the DOPO neuron in the parameter space of $P$ and $I_{\text{ext}}$. The color map represents spiking frequencies calculated by numerical simulations. Red and blue dots obtained by linear stability analysis represent points where the Andronov-Hopf (AH) bifurcation and saddle-node bifurcation on a limit cycle (SNLC) occur, so they characterize class-II and class-I neurons, respectively. **c** Experimental results of tomographic measurement of spiking frequencies along lines A, B, and C in (**b**). The color map represents Fourier signals of time evolutions of DOPO amplitudes. Cyan points are firing rates estimated by adding up the number of spikes.

## Spontaneous modification of collective dynamics

We next investigated the synchronization phenomenon of networked DOPO neurons, which is an essential factor in signal processing of the SNN. Networks of 60 DOPO neurons were constructed as depicted in Fig. 3a. In each network, 15 neurons form an all-to-all connected cluster (like the Kuramoto model[38,39]; also see method), and four such clusters are sparsely connected. This network structure was encoded into connections of both $v$- and $w$-DOPOs (i.e., $\gamma = \gamma' \equiv J_k$ ). Four independent sets of such ensembles, consisting of 60 DOPO neurons with different coupling strengths ($\widetilde{J_k}$ = 0, 0.025, 0.05, and 0.075), were implemented. We set $I_{\text{ext}} = 0$ and used uniform $i$-independent coupling $|J_{vw}| = |J_{wv}|$. Then, $i$-dependence of the pump was tuned to control the distribution of $\omega_i(P_i)$, and were assigned to four 15-neuron clusters labeled A to D in descending order of firing rate. The distribution of firing rates of the 60 DOPO neurons is shown in Fig. 3b. Without coupling ($J_k$= 0), the firing rates are widely spread by applied pump $P_i$. With weak coupling at $\widetilde{J_k}$ = 0.025, the distribution of firing rates is shown as four uniform frequencies, suggesting that coupled DOPO neurons show obvious synchronization in each 15-neuron cluster; however, the four clusters were unsynchronized. As $J_k$ increases, mean and variance of the firing rate are decreased. Figure 3c shows the time evolutions of DOPO neurons with coupling at $\widetilde{J_k}$ = 0.075, where $\theta_i \equiv \arg(v_i + i\, w_i)$ is a phase defined on the $v$ - $w$ plane of amplitudes of paired DOPOs (see Supplementary S-E for the other coupling strengths $\widetilde{J_k}$). Figure 3d shows the phase change $\Delta\theta_i$ in each four cavity circulations. We evaluated the order parameter $r$ defined as $r = \left|\frac{1}{N}\sum_j e^{i\theta_j}\right|$ for each cluster ($N$ = 15) and for all the neurons ($N$ = 60) as shown in Fig. 3e. By increasing coupling strength to $\widetilde{J_k}$=0.075, the four clusters became intermittent synchronization and the order parameter occasionally reached $r \sim 1$.

Such behavior of the order parameter can be understood from the Kuramoto model[38,39]. However, it should be noted that firing rates are significantly decreased from their original values as $J_k$ increases as shown in Fig. 3b, which is different from the standard behavior in the Kuramoto model. Further analysis suggests that the change in firing rate is induced by an effective change in spiking mode of synchronized DOPO neurons as an ensemble. The dynamics of the networked DOPO neurons defined by equations (1) and (2) can be rewritten with $\sqrt{R_i}e^{i\theta_i} \equiv v_i + i\, w_i$ as

$$\frac{d\theta_i}{dt} = \omega_0 - J_k \sum_{j\neq i} \varepsilon_{ij} \sin(\theta_i - \theta_j) + \frac{R_i}{4} \sin 4\theta_i \,, \tag{5}$$

$$\frac{dR_i}{dt} = 2P_i R_i + 2J_k R_i \sum_{j\neq i} \varepsilon_{ij} \cos(\theta_i - \theta_j) - \frac{R_i^2}{2} (\cos 4\theta_i + 3) \,, \tag{6}$$

where $\varepsilon_{ij} = \sqrt{R_j/R_i}$. The quadrature components of the DOPOs are neglected here. Equation (5) is analogous to the standard Kuramoto model (see Method). Equation (6) indicates that the pump term should be renormalized as $P_i' = P_i + J_k \sum_{j\neq i} \varepsilon_{ij} \cos(\theta_i - \theta_j)$, and the second term on the right-hand side can be reduced to $r(N-1)J_k$ under the approximation $\varepsilon_{ij} \sim 1$. This result suggests that the synchronization with a large order parameter ($r \sim 1$) increases the pump term effectively and causes the spontaneous change in the spiking mode from class-II to class-I. The spiking dynamics of the class-I neuron has an escape time at proximity of unstable stationary points ($\Delta\theta_i \sim 0$) and the timing of the next spike is flexibly tuned at these points (see Fig. 3d, in which the width of the blue areas with $\Delta\theta_i \sim 0$ is frequently changed). It was also found that during a long escape time at proximity of unstable stationary points, order parameter $r$ tends to increase up to 1, as shown in Fig. 3e, suggesting that the timing of spikes is tuned spontaneously to increase the order parameter. For the unsynchronized ensemble, on the other hand, the effective change in the pump term is suppressed with a small $r$, and the spiking mode reverts to class-II. Since there is no stationary point in the spiking dynamics of the class-II neuron, the tuning of the spike timing is suppressed. Consequently, the self-tuning effect of collective spiking dynamics of clustered neurons assists the overall synchronization, even though each cluster has significantly different firing rates.

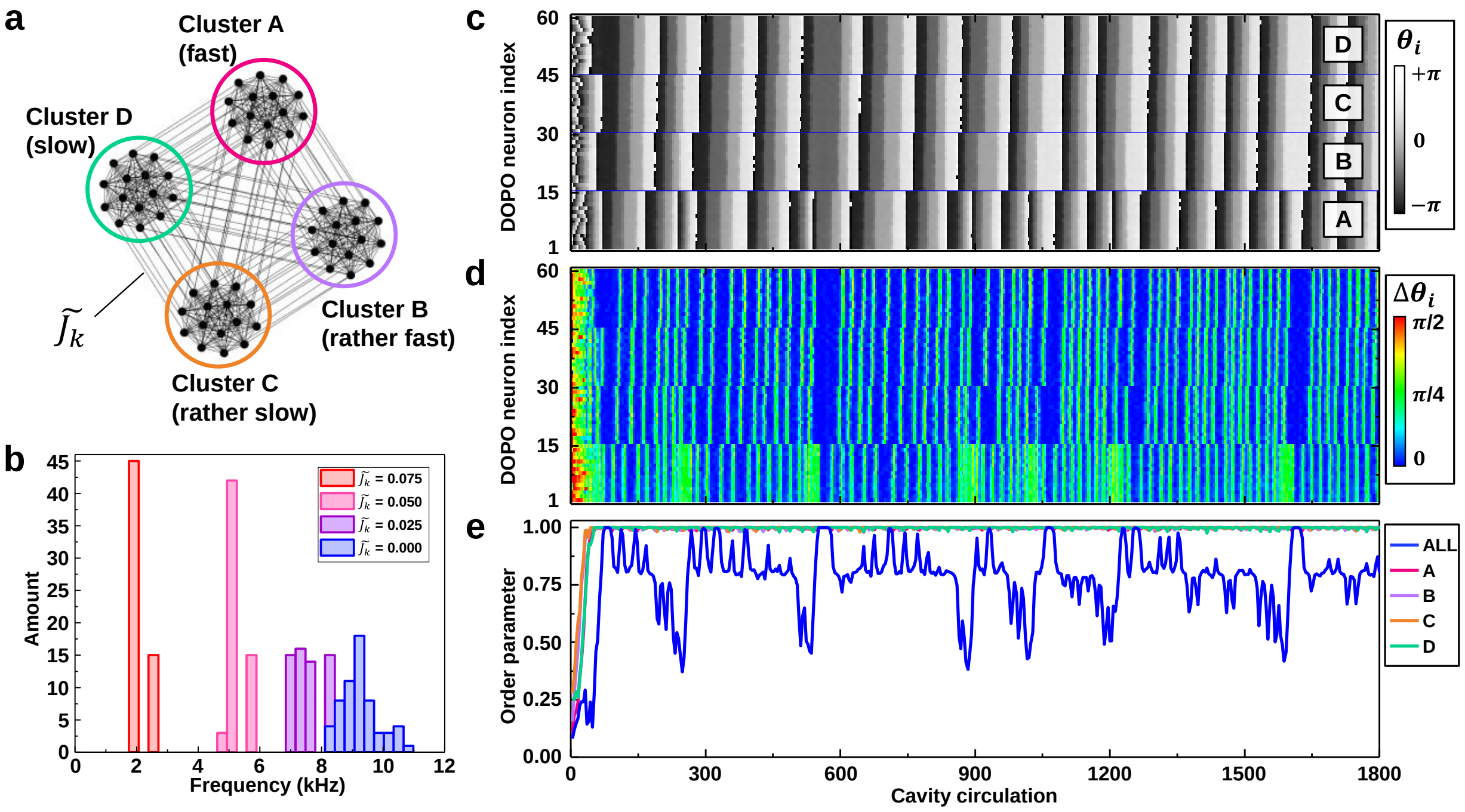

**Fig. 3** Synchronization experiment of clustered Kuramoto models. **a** Structure of DOPO-neuron network consisting of four clusters of 15 neurons. **b** Firing-rate distribution of 60 neurons. For $J_k = 0$, the distribution is spread as designed on the basis of pump amplitude. Coupling $J_k$ causes the distribution to shrink and shift toward lower frequency. **c** Time evolutions of phase $\theta_i$ of the $i$th DOPO neuron with coupling of $\widetilde{J_k}$ = 0.075, where $\theta_i$ is an angle defined in the $v$-$w$ plane of the coupled DOPOs. **d** Phase change $\Delta\theta_i$ per four cavity circulations. **e** Order parameter $r$ for each cluster and for all 60 neurons. Synchronization can be characterized by $r \sim 1$.

## Combinatorial optimization using self-tuning collective dynamics

To understand the effect of this spontaneous change in firing rates of local clusters on the whole network dynamics, an analogy with complex frustrated spin system, which is named Ising model[40], is considered as follows. The state of Ising model is described in terms of Ising energy given by $E_{\text{Ising}} = -\sum_{i<j} J_{ij}\sigma_i\sigma_j$, where $\sigma_i = \{-1, +1\}$ denotes the $i$th Ising spin state, and $J_{ij}$ is a symmetric spin-spin interaction matrix between the $i$th and $j$th spins. The Ising spin state can be represented by the binary phase state of the $v$-DOPO. The spin-spin interaction $J_{ij}$ is implemented only for synaptic connections between $v$-DOPOs ($\gamma = -J_k$, $\gamma' = 0$). The external bias $I_{\text{ext}}$ is set to zero. Of particular interest is relaxation of the DOPO SNN to configurations with lower Ising energy, which can be used to solve many combinatorial optimization problems. An instance was solved as the first benchmark problem: a highly frustrated network of 150 Ising spins coupled by symmetric connections with edge density of 50%. This instance has been investigated by using various algorithms, such as the one used in a spin-glass server[41] and physical systems based on networked DOPOs, such as a coherent Ising machine (CIM)[42]. The best-known solution of the instance has an $E_{\mathrm{Ising}}$ of $-700$. Note that the probability of reaching the best-known solution by using the CIM was less than 1%, suggesting that this instance is rather hard to solve. The time-dependent and node-independent pump $P_i = P_0(t)$ was applied to the DOPOs, and the amplitude linearly increased with time. Time evolution of measured $v$-DOPO amplitudes with coupling strength $\widetilde{J_k}$ = 0.250 is shown in Fig. 4a. As a reference, dynamics of uncoupled DOPO neurons is shown in the top panel. This figure suggests that firing rate gradually decreases with time, and the spiking dynamics terminates at the end of calculation because the pump $P_0(t)$ finally reaches a value exceeding the spiking region. When the network is connected (see lower panel), DOPO neurons show highly irregular and complex spiking behavior. Time evolutions of the Ising energy with various coupling strength ($\widetilde{J_k}$ = 0.083, 0.167, and 0.250) are shown in Fig. 4b. Lower-energy solutions were found with higher coupling strengths at $\widetilde{J_k} = 0.250$, and the best-known solution was obtained with success probability of over 10%. To understand the dependence of the dynamics on coupling strength, the relationship between total firing count of each DOPO neuron and local energy $E_{\mathrm{loc},i}$ of the final solution was analyzed as shown in Fig. 4c. The local energy is defined as $E_{\mathrm{loc},i} = -\sum_j J_{ij}\sigma_i\sigma_j$, which is related to Ising energy by $E_{\mathrm{Ising}} = \frac{1}{2}\sum_i E_{\mathrm{loc},i}$. With increasing coupling strength, a positive correlation between the firing count and local energy appears. As clarified in the above discussion, synchronization of DOPO neurons causes firing rates to change. The order parameter can be related to local energy $E_{\mathrm{loc},i}$ by using certain approximations (see Supplementary S-F); thus, renormalized pump amplitude can be rewritten as $P_i = P_0(t) - \frac{1}{2}\tilde{J}_k E_{\mathrm{loc},i}(t)$. This means that DOPO neurons with higher $E_{\mathrm{loc},i}$ (energetically unstable nodes) show higher firing rates and those with lower $E_{\mathrm{loc},i}$ (energetically stable nodes) show lower firing rates. Consequently, the DOPO neural network spontaneously tries to flip Ising spins frequently, primarily on energetically unstable nodes, and such a selective spin-flip mechanism might be a key factor in accelerating the relaxation to lower energy states. This behavior of the DOPO neural network is similar to that of the state-of-the-art algorithms for combinatorial optimization such as extremal optimization[43] and methods related to self-organized criticality[44].

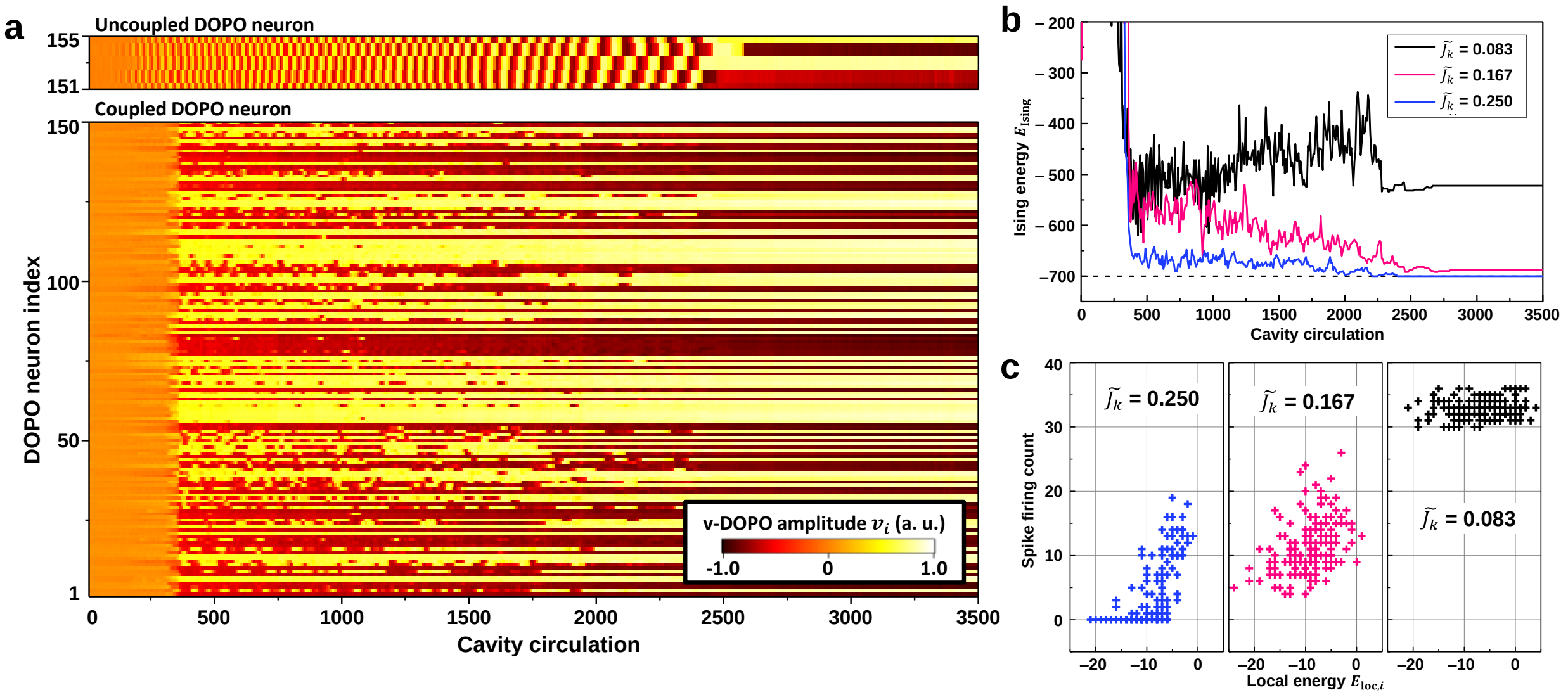

**Fig. 4** Spiking dynamics when solving a 150-node Ising problem. **a** Time evolutions of $v$-DOPO amplitudes with couplings of $\tilde{J}_k$ = 0.250 (lower) and without coupling (top). Pump amplitudes increased linearly with time. **b** Time evolutions of Ising energy for $\tilde{J}_k$ = 0.083, 0.167, and 0.250. Strong couplings give the best-known solution. **c** Relationship between local energy and firing rates. A positive correlation between firing rate and local energy can be found with strong couplings, suggesting that energetically unstable neurons have high firing rate due to the self-tuning effect of the spiking mode.

**Discussion**

We confirmed that spiking dynamics of the DOPO neurons could be controlled from the class-II to class-I neuronal mode by increasing the optical pump amplitude, and the firing rate was modulated dynamically due to the crossover of different spiking modes. This flexible controllability of spiking modes induced a self-tuning effect of collective dynamics of the DOPO neurons. Because the pump amplitudes of clustered DOPO neurons can be approximately renormalized as $P_i' = P_i + r(N-1)J_k$, the pump amplitude was effectively increased as the order parameter $r$ of synchronization was increased. Our experimental results showed that the firing rates of neurons were modulated reflecting the order parameter on the present situation. Such spontaneous modification of collective dynamics assisted the synchronization even though neurons have significantly different firing rates. Additionally, we utilized the self-tuning effect of the firing rates to improve the optimization process of the Ising spin network. Since the renormalized pump amplitude can be rewritten as $P_i' = P_i - \frac{1}{2}J_k E_{\mathrm{loc},i}$ in the antiferromagnetic spin network, the firing rates of neurons were modulated according to the local energy $E_{\mathrm{loc},i}$. The firing-rate selectivity for the local energy might be an effective way to find global lower-energy solutions. The present DOPO neural network inherently includes such a dynamical optimization process thanks to the self-tuning effect of the collective spiking dynamics.

The spiking modes and the firing rates of DOPO neurons were tuned adaptively depending on the network structure, coupling strength, and synchronization with adjacent neurons. We expect that this characteristic will provide an additional degree of freedom for designing information processing based on artificial SNNs.

## Method

### Analysis of linear stability

Hereafter, the bifurcation of a single neuron is discussed without consideration of interneuron couplings and external fields ($\gamma = \gamma' = I_{\text{ext}} = 0$). It is assumed that the $i$-dependence can be neglected in the notations. Analysis of linear stability based on equations (1) and (2) explains what kinds of bifurcations appear. A linearized form around an equilibrium point $\begin{pmatrix} v \\ w \end{pmatrix} = \begin{pmatrix} v_e \\ w_e \end{pmatrix}$ is given by $\frac{d}{dt}\begin{pmatrix} v \\ w \end{pmatrix} = M\begin{pmatrix} v \\ w \end{pmatrix}$ with $M = \begin{pmatrix} P - 3v_e^2 & -\omega_0 \\ \omega_0 & P - 3w_e^2 \end{pmatrix}$. For small $P(> 0)$, equilibrium found at $\begin{pmatrix} v_e \\ w_e \end{pmatrix} = \begin{pmatrix} 0 \\ 0 \end{pmatrix}$ has eigenvalues of $M$ given by $\lambda = P \pm i\omega_0$, suggesting an AH bifurcation. For large $P$, equilibriums can be found at the tangency points of nullclines around $\begin{pmatrix} v_e \\ w_e \end{pmatrix} \sim \begin{pmatrix} \pm\sqrt{P/3} \\ \mp 2P\sqrt{P}/3\omega_0 \end{pmatrix}$. Around such equilibriums, two of $\lambda$ can be real and positive values that characterize the SNLC bifurcation. Using numerical calculations, is possible to precisely estimate equilibriums and corresponding c, and the points where AH or SNLC bifurcations occur can then be evaluated, as shown in Fig. 2b. In addition, by simulating the dynamics based on equations (1) and (2), it is also possible to calculate spiking frequency. These two kinds of calculations are consistent with each other. See Supplementary S-A and S-B for more mathematical explanations.

### Spiking frequency

From equations (5) and (6) (in which $J_k = 0$ is set and $i$-dependence is neglected), spiking frequency $\omega(P)$ can be approximately obtained. $R$ can be roughly evaluated under the condition $\frac{dR}{dt} = 0$ as $R \sim \frac{2P}{\pi}\int_{-\pi}^{\pi} \frac{d\theta}{3+\cos 4\theta} = \sqrt{2}P$, meaning that the number of photons ($\propto R = v^2 + w^2$) increases linearly with pump amplitude. From this relation $\frac{d\theta}{dt} = \omega_0 + \frac{P}{\sqrt{8}} \sin 4\theta$ is obtained, and the period of the oscillator can be approximately evaluated as $T = \int_{-\pi}^{\pi} \frac{d\theta}{\omega_0 + \frac{P}{\sqrt{8}} \sin 4\theta} = \frac{2\pi}{\sqrt{\omega_0^2 - P^2/8}}$. Finally, we obtain $\omega(P)$ in equation (3). See Supplementary S-C for more detail about this calculation.

**Kuramoto model**

The Kuramoto model, which provides a paradigm for understanding the mechanism of synchronization phenomena of coupled nonlinear oscillators[38,39], is briefly introduced hereafter. The Kuramoto model for *N* nonlinear oscillators with all-to-all coupling of strength $J_k$ is given as

$$\frac{d\theta_i}{dt} = \omega_i - J_k \sum_j \sin(\theta_i - \theta_j)\ , \qquad (7)$$

where $\theta_i$ and $\omega_i$ are the phase and natural angular frequency of the *i*th nonlinear oscillator for *i* = 1, 2, …, *N*. The synchronization of coupled oscillators can be understood as a kind of phase transitions characterized by order parameter $r$ defined as $r = \frac{1}{N} \sum_j e^{i\theta_j}$. Under the assumption that $\omega_i$ has a distribution with variance $\sigma_\omega$, $r$ is zero for small $J_k/\sigma_\omega$ and becomes finite for large $J_k/\sigma_\omega$ (also see Supplementary S-D). In our experiments, the distribution of $\omega_i$ was introduced by controlling pump amplitude $P_i$ through equation (3).

## Acknowledgments

This research was funded by the Impulsing Paradigm Change through Disruptive Technologies (ImPACT) Program of the Council of Science, Technology and Innovation (Cabinet Office, Government of Japan). K.A. is partially supported by Japan Agency for Medical Research and Development (AMED) under Grant Number JP20dm0307009. The authors thank Hiroyuki Tamura for his support during this research.

## Author contributions

K.I., T.I., and T.L. proposed the project. T.I. performed the experiments. K.I. performed data analysis and numerical simulations. T.L. and K.A. supported theoretical analysis of computational neuroscience. T.I., K.I., T. Ikuta, T.H., and H.T. contributed to building the DOPO network system. K.E., T.U., and R.K. contributed to building the PPLN modules. T.I., K.I., T.L., K.A., and H.T. wrote the manuscript with inputs from all authors.

## Data availability

The data that support the findings of this study are available from the corresponding author upon reasonable request.

## Competing financial interests

The authors declare no competing financial interests.

Supplementary Materials for

# Collective and synchronous dynamics of photonic spiking neurons

Takahiro Inagaki*, Kensuke Inaba*, Timothée Leleu, Toshimori Honjo, Takuya Ikuta, Koji Enbutsu, Takeshi Umeki, Ryoichi Kasahara, Kazuyuki Aihara, and Hiroki Takesue

*Correspondence to: Takahiro Inagaki (takahiro.inagaki.vn@hco.ntt.co.jp) and Kensuke Inaba (kensuke.inaba.yg@hco.ntt.co.jp)

**This PDF file includes:**

Supplementary Text: 6 sections with 5 figures

## Section S-A: Numerical simulation

In this section, we show numerical simulation results for investigating the dynamics of the single DOPO neuron based on the following coupled ordinary differential equations (ODEs):

$$\frac{dv}{dt} = -v + (1+P)v^* - |v|^2 v + J_{vw}\mathrm{Re}(w) + I_{ext}, \quad \text{(A1)}$$

$$\frac{dw}{dt} = -w + (1+P)w^* - |w|^2 w + J_{wv}\mathrm{Re}(v), \quad \text{(A2)}$$

where variables $v$ and $w$ are amplitudes of DOPOs called v- and w-DOPO, $v^*$ and $w^*$ are their complex conjugates, $\mathrm{Re}(x)$ and $\mathrm{Im}(x)$ are in-phase and quadrature-phase components of the DOPO amplitudes for $x = v, w$. Parameters $P$ and $I_{ext}$ are the optical amplitude of the pump and the external bias term, and $J_{vw}$ and $J_{wv}$ are connections between $v$- and $w$-DOPOs, where the coupling is implemented by the measurement feedback method and is applied only to the in-phase component. Without coupling, each equation is the same as that describing the dynamics of a single DOPO[1-3]. For $P > 0$, the real components of $v$ and $w$ are amplified, while the imaginary ones are damped, which means that we can neglect the latter components. Note that our definition of the pump amplitude $P$ is slightly modified from those used in the previous papers, and the use of $P$ is convenient for discussing the spiking dynamics found in the real components. The previous definition $p$ is an optical amplitude normalized by the threshold value of an uncoupled DOPO, and $P$ is given by $p-1$. Antisymmetric coupling with $J_{vw} = -J_{wv}$ (generally speaking, asymmetric coupling with $J_{vw}J_{wv} < 0$) induces energy transport between $v$ and $w$ and yields a stable limit cycle, which can be interpreted as the repetitive spiking process. Equations (A1) and (A2) can be rewritten as

$$\frac{d\mathrm{Re}(v)}{dt} = P\mathrm{Re}(v) - |v|^2\mathrm{Re}(v) + J_{vw}\mathrm{Re}(w) + I_{ext} \ , \tag{A3}$$

$$\frac{d\mathrm{Re}(w)}{dt} = P\mathrm{Re}(w) - |w|^2\mathrm{Re}(w) + J_{wv}\mathrm{Re}(v) \ , \tag{A4}$$

$$\frac{d\mathrm{Im}(v)}{dt} = (-2-P)\mathrm{Im}(v) - |v|^2\mathrm{Im}(v) \ , \tag{A5}$$

$$\frac{d\mathrm{Im}(w)}{dt} = (-2-P)\mathrm{Im}(w) - |w|^2\mathrm{Im}(w) \ , \tag{A6}$$

where Eqs. (A3) and (A4) are the forms corresponding to Eqs. (1) and (2) in the main text.

We here define two parameters that characterize the properties of bifurcations (see also section S-B): a basic frequency $\omega_0 \equiv \sqrt{-J_{vw}J_{wv}}$ and an anisotropy of v- and w-DOPOs $\alpha \equiv \left|\frac{J_{wv}}{J_{vw}}\right|$, which are controllable in the present DOPO system. In the current study we mainly considered the isotropic case $\alpha = 1$. It is convenient to use $\omega_0$ as the time unit in the ODEs

Eqs. (A1) and (A2) so that we can rescale the above equations to dimensionless forms by using $\tilde{v} = v/\sqrt{\omega_0}$, $\tilde{w} = w/\sqrt{\alpha\omega_0}$, $\tilde{P} = P/\omega_0$, $\tilde{I}_{ext} = I_{ext}/\sqrt{\omega_0}^3$, and $\tilde{t} = t\omega_0$. We then obtain the following dimensionless forms: $\frac{d\tilde{v}}{d\tilde{t}} = -\tilde{v} + (1+\tilde{P})\tilde{v}^* - |\tilde{v}|^2\tilde{v} \mp \mathrm{Re}(\tilde{w}) + \tilde{I}_{ext}$, and $\frac{d\tilde{w}}{d\tilde{t}} = -\tilde{w} + (1+\tilde{P})\tilde{w}^* - \alpha|\tilde{w}|^2\tilde{w} \pm \mathrm{Re}(\tilde{v})$.

In the present numerical simulation, we take into account the noise terms given by $\sqrt{|v|^2+\frac{1}{2}}dW_{\mathrm{Re}(v)}$, $\sqrt{|v|^2+\frac{1}{2}}dW_{\mathrm{Im}(v)}$, $\sqrt{|w|^2+\frac{1}{2}}dW_{\mathrm{Re}(w)}$ and $\sqrt{|w|^2+\frac{1}{2}}dW_{\mathrm{Im}(w)}$, where $dW_x$ is the independent Gaussian noise for real and imaginary parts of $v$- and $w$-DOPOs[1-3]. Except for the following numerical simulation, we neglect these noise terms for simplicity and to focus on the mathematical study of the bifurcations.

Figure A1 shows numerically simulated time series of DOPO amplitudes and the corresponding trajectories in the v-w space with the external bias $I_{ext} = 0$ or $\tilde{I}_{ext} = -0.3$ for $\tilde{P} = 0.3, 1.1, 1.8$ and $2.5$. Numerical simulations were calculated using the Runge–Kutta method with parameters $\mathrm{dt} = 0.05$ and $dW_x \equiv 0.02\times$Gaussian noise. For comparison, trajectories without noise terms ($dW_x = 0$) are also presented in the figure. We found that a firing rate decreases as the pump amplitude increases, and qualitative features of the time series and the trajectories are also changed. For a small $P$, we can find a simple limit cycle around $(\mathrm{v}, \mathrm{w}) = (0,0)$ that corresponds to the vacuum state. Here, eigenvalues of the Jacobian matrix are given by $\lambda_{\pm} \sim \tilde{P} \pm i$. Thus, we can state that the Andronov-Hopf (AH) bifurcation occurs at around $\tilde{P} = 0$. Figure A1 clearly shows that the two nullclines become tangential for a large $\tilde{P}$, suggesting that the saddle-node bifurcation on the limit cycle (SNLC) emerges.

With Fig. A1, we provide the details of the linear stability analysis shown in Method. A linearized form around an equilibrium point $\begin{pmatrix} v \\ w \end{pmatrix} = \begin{pmatrix} v_e \\ w_e \end{pmatrix}$ is given by $\frac{d}{dt}\begin{pmatrix} v \\ w \end{pmatrix} = M\begin{pmatrix} v \\ w \end{pmatrix}$ with $M = \begin{pmatrix} \frac{\partial f(v,w)}{\partial v} & \frac{\partial f(v,w)}{\partial w} \\ \frac{\partial g(v,w)}{\partial v} & \frac{\partial g(v,w)}{\partial w} \end{pmatrix} = \begin{pmatrix} P - 3v_e^2 & -\omega_0 \\ \omega_0 & P - 3w_e^2 \end{pmatrix}$, where a function $f(v,w)$ $[g(v,w)]$ is right hand of (A3) [(A4)]. The bottom panels of Fig. A1 show that equilibrium at around $\begin{pmatrix} v_e \\ w_e \end{pmatrix} = \begin{pmatrix} 0 \\ 0 \end{pmatrix}$ is found for small $P(> 0)$. Around this equilibrium, eigenvalues of $M$ are given by $\lambda = P \pm i\omega_0$, suggesting an AH bifurcation. The top panels of Fig. A1 show that equilibriums can be found at the tangency points of nullclines around $\begin{pmatrix} v_e \\ w_e \end{pmatrix} \sim \begin{pmatrix} \pm\sqrt{P/3} \\ \mp 2P\sqrt{P}/3\omega_0 \end{pmatrix}$ for large $P$. Around such equilibriums, two of $\lambda$ can be real and positive values that characterize the SNLC bifurcation. Using numerical calculations, it is possible to precisely estimate equilibriums and corresponding $v_e$ and $w_e$, and the points where AH or SNLC bifurcations occur can then be evaluated, as shown in Fig. 2b in the main text. Numerical simulations also showed that, at the proximity of cross points where these two bifurcation sets merge with each other, other kinds of bifurcations, namely codimension-two bifurcations, occur but are not detailed here. In addition, by simulating the dynamics based on equations (1) and (2) in the main text, it is also possible to calculate spiking frequency [color map of Fig. 2b]. It is confirmed that these two kinds of calculations are consistent with each other: Namely, the region surrounded by red (AH) and blue (SNLC) curves coincides with the colored region with finite values of spiking frequency.

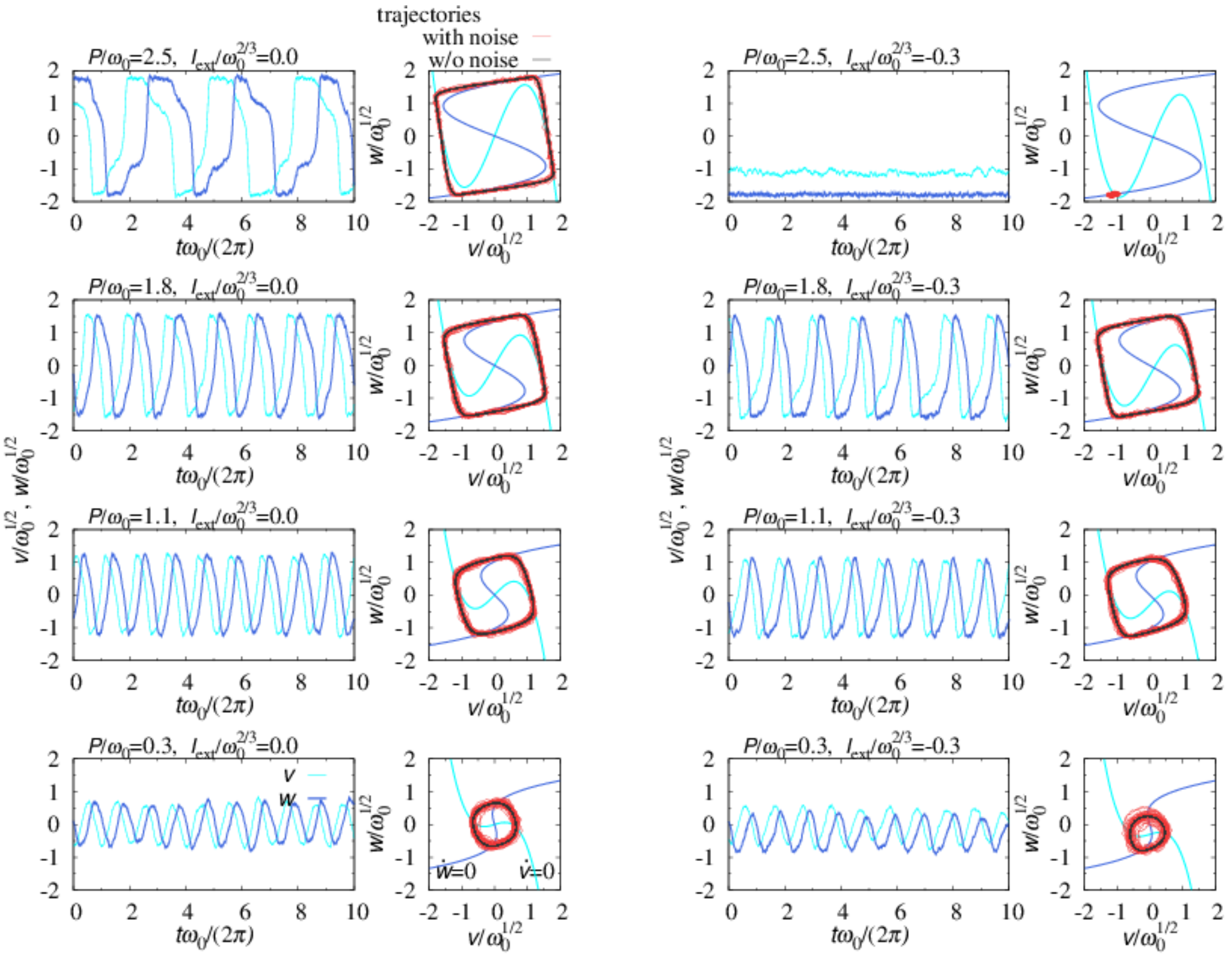

**Fig. A1** Time series of the DOPO amplitudes, and the trajectories (red) in the v-w space and the nullclines, which are curves satisfying dv/dt=0 (cyan) and dw/dt=0 (blue). For comparison, trajectories calculated without noise terms are also shown (black), which correspond to the orbit of the limit cycles.

## Section S-B: Bifurcation model

Here, we analytically discuss bifurcation models of the present system. In what follows, to capture the essence of the bifurcations found in the present experiments, we do not take into account noise terms and quadrature components of DOPOs, and thus variables $v$ and $w$ represent the in-phase components. We do not consider the external bias term, either, as the first step. We now consider the following equations:

$$\frac{dv}{dt} = Pv - (v^2 + \beta w^2)v + J_{vw}w, \tag{B1}$$

$$\frac{dw}{dt} = Pw - (\beta v^2 + w^2)w + J_{wv}v, \tag{B2}$$

where, for convenience, we introduce a new parameter $\beta$ that characterizes the property of nonlinear terms. Although this parameter is zero ($\beta = 0$) and is uncontrollable in the present experiments, it helps us to discuss the change of types of bifurcations. We should note that some physical systems can be described by the analogical equation to the above with $\beta = 1$ [46], and thus this parameter $\beta$ allows us to clarify the difference of the bifurcation mechanisms of different physical systems.

Here we assume $J_{vw}J_{wv} < 0$ and consider only the case of $J_{vw} < 0$ and $J_{wv} > 0$. The opposite-sign case can be extracted from the same discussions below under the transformation of $v \to -v$. We rescale $w$ as $\sqrt{\alpha}w$, and then we can rewrite the above equations using $z = v + iw$ as

$$\frac{dz}{dt} = i\omega_0 z + Pz - \frac{1+\alpha}{2}\left(|z|^2 z - \frac{1-\beta}{4}(z^2 - z^{*2})z^*\right)$$

$$-\frac{1-\alpha}{2}\left(|z|^2 z^* + \frac{1+\beta}{4}(z^2 - z^{*2})z\right), \qquad \text{(B3)}$$

which can be rewritten with $z = \sqrt{R}e^{i\theta}$ and $R = v^2 + w^2/\alpha$ as

$$\frac{d\theta}{dt} = \omega_0 + R\frac{1-\beta}{4}\left(\frac{1+\alpha}{2}\sin 4\theta - \frac{1-\alpha}{2}2\sin 2\theta\right), \qquad \text{(B4)}$$

$$\frac{dR}{dt} = 2PR - 2R^2\left[\frac{1+\alpha}{2}\left(\frac{3+\beta}{4} + \frac{1-\beta}{4}\cos 4\theta\right) + \frac{1-\alpha}{2}\cos 2\theta\right]. \qquad \text{(B5)}$$

From the above expressions we can easily understand that $\omega_0$ corresponds to the angular frequency of the oscillator at $\beta = 1$, or at the limit of $R \to 0$ corresponding to the $P \to +0$ limit, where we have a simple oscillator with $\frac{d\theta}{dt} = \omega_0$. We can thus expect that the AH bifurcation should occur near these two limits. For example, for $\alpha = \beta = 1$, we can obtain the normal form of the AH bifurcation: $\frac{dz}{dt} = i\omega_0 z + Pz - |z|^2 z$. From another viewpoint, we obtain two separable variables ($\theta$ and $R$): $\frac{d\theta}{dt} = \omega_0$ and $\frac{dR}{dt} = 2PR - 2R^2$. However, for $\beta = 0$ and $P > 0$, the connection between $\theta$ and $R$ induces change of bifurcations from AH to SNLC types. Roughly speaking, we can derive the normal form of the saddle-node bifurcations, too: For example, at $\alpha = 1$ and $\beta = 0$ the form $\frac{d\theta}{dt} = \omega_0 + \frac{R}{4}\sin 4\theta$ can be reduced to $\frac{d\theta}{dt} = \omega_0 +$ $\frac{P}{\sqrt{8}}(-1+8\theta^2)$. This point will be discussed again in section S-C. We can thus understand that parameters $P$ and $\beta$ play important roles in the mechanism of the bifurcations.

## Section S-C: Mathematical expressions

We derive some mathematical expressions to estimate spiking frequencies and bifurcation points. Such mathematical expressions help us compare the experimental results with theory and numerical calculations. We should note that in the experiments all of the parameters are influenced by various noises, losses, and fluctuations, and thus it is difficult to determine the magnitude of parameters precisely. Therefore it is important to use mathematical expressions to discuss the quantitative aspects of the experiments.

*Spiking frequency.* In the main text, we used parameters $\alpha = 1$ and $\beta = 0$. Here we obtain $\frac{d\theta}{dt} = \omega_0 + \frac{R}{4}\sin 4\theta$ and $\frac{dR}{dt} = 2PR - \frac{R^2}{2}(3 + \cos 4\theta)$ . As discussed above, at the limit of $P \to +0$, a firing frequency $\omega$ is equivalent with the basic frequency $\omega_0 (\equiv \sqrt{-J_{vw}J_{wv}})$. The experimental and simulation results suggest that the firing rate decreases as $P$ increases. In the following, we analyze $\omega$ as a function of pump amplitude $P$. We first estimate an averaged $R$ in one cycle of oscillations (during a change of $\theta$ in $[-\pi, \pi]$), which corresponds to the static component of $R$ with a condition of $\frac{dR}{dt} = 0$. It can be estimated as $R_{\text{ave}} = \frac{2P}{\pi}\int_{-\pi}^{\pi} \frac{d\theta}{3+\cos 4\theta} = \sqrt{2}P$. In our optical system, it means that the number of photons ($\propto R = v^2 + w^2$) increases linearly with increasing pump amplitudes $P$. By neglecting the oscillatory component of $R$, we obtain $\frac{d\theta}{dt} = \omega_0 + \frac{P}{\sqrt{8}}\sin 4\theta$. Then we can approximately evaluate the period of the oscillator $T = \int_{-\pi}^{\pi} \frac{d\theta}{\omega_0 + \frac{P}{\sqrt{8}}\sin 4\theta} = \frac{2\pi}{\sqrt{\omega_0^2 - P^2/8}}$ and also the angular frequency $\omega$, corresponding to the firing rate, as a function of $P$ as follows:

$$\omega(P) = \omega_0\sqrt{1 - \frac{P^2}{8\omega_0^2}}. \tag{C1}$$

The limit of $P \to +0$ leads the basic frequency $\omega_0$ as mentioned above, suggesting that the AH bifurcation (class II neuron) can be found near $P \sim 0$.

At $P = \sqrt{8}\omega_0$, the frequency gradually reaches zero, suggesting the SNLC bifurcation (the class I neuron). This point $P_{SNLC}$ is exactly the same as the analytically obtained expression of the bifurcation point as mentioned below. Near $P \sim P_{SNLC}$, four points of $\theta$ with $\frac{1}{4}(\frac{\pi}{2} + 2n\pi)$ satisfy $\frac{d\theta}{dt} \sim 0$, suggesting that $\theta$ stays near these points in almost all time of one cycle. Then we obtain the period of the saddle node bifurcations $T \sim \frac{4}{\omega_0}\int_{-\infty}^{\infty} \frac{d\theta}{1 - \frac{P}{P_{SLNC}} + \frac{8P}{P_{SLNC}}\theta^2}$. Finally, we obtain $\omega_{P \sim P_{SNLC}}(P) = \omega_0\sqrt{\frac{2P}{P_{SNLC}}}\sqrt{1 - \frac{P}{P_{SNLC}}}$, which is equivalent with the above expression near $P \sim \sqrt{8}\omega_0 = P_{SNLC}$. Note that there are four points satisfying $\frac{d\theta}{dt} \sim 0$ for $\alpha = 1$ and $I_{ext} = 0$ because of two kinds of symmetry, sign-inversion $[(v, w) \to (-v, -w)]$ and $v$-$w$-inversion $[(v, w) \to (w, v)]$ symmetry: For $\alpha \neq 1$ the latter is broken, and for $I_{ext} \neq 0$ both are broken.

We compare the above approximated expressions with the numerical calculations. Figure C1 shows spiking frequencies as a function of the pump amplitude $P$. Near both end points ($P \sim 0$ or $P \sim P_{SNLC}$), the above expression $\omega(P)$ agrees well with the numerical results. We should note that the numerical simulations have been done with a small enough $dt$ to reproduce the continuous limit. We found that discreteness of $dt$ induces quantitative deviations. However, simulations with a larger $dt$ and experiments with discrete timing of feedback signals show qualitatively similar behavior. We should note that the nonlinear terms such as $Pv - v^3$, which are optically implemented in nonlinear optics, are perfectly continuous in experiments, even though the linear terms implemented by the measurement feedback method are discrete.

*Bifurcation points.* We next show the mathematical expressions of the bifurcation points. Without external fields $\tilde{I}_{ext} = 0$, we can calculate analytically the $P$ values where the bifurcation occurs. As mentioned above, the $P \to 0$ limit reproduces the normal form of the AH bifurcation. From this point, we can obtain the bifurcation point of $\tilde{P}_{\mathrm{AH}} = 0$. For the SNLC bifurcations, we can calculate the points where two nullclines have tangency. Thus, we can easily calculate the bifurcation point $\tilde{P}_{SNLC}$ by solving a cubic equation $4\alpha\left(4+\tilde{P}^2\right)^3 = 27\tilde{P}^4(\alpha+1)^2$, which yields

$$\tilde{P}_{SNLC} = \sqrt{\frac{1}{2} + \frac{9}{4}A + \frac{3}{2}\sqrt{(-14+9A)(2+A)}\cos\Psi}, \qquad \text{(C2)}$$

where $\Psi = \frac{1}{3}\mathrm{acos}\left(\frac{(27A^2-36A-52)}{(-14+9A)^{\frac{3}{2}}(2+A)^{\frac{1}{2}}}\right) + \frac{4\pi}{3}$ and $A = \alpha + \frac{1}{\alpha}$. For $\alpha = 1$, we obtain $\tilde{P}_{SNLC} = \sqrt{8}$ as mentioned above.

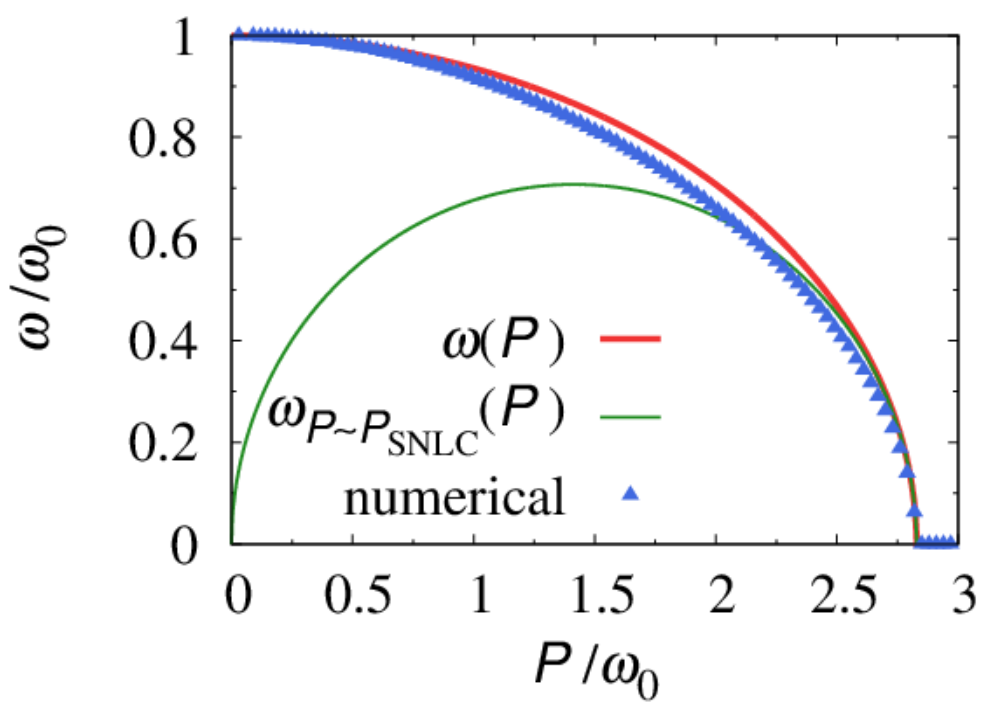

**Fig. C1** Spiking frequency as a function of pump amplitude $P$ calculated by the numerical simulations (blue triangles) and analytical forms (red and green lines).

## Section S- D: Spontaneous modification of spiking frequency

We discuss the connected spiking DOPO neurons. Here, we set the parameters $\alpha = 1$, $\beta = 0$, and $I_{\text{ext}} = 0$ for simplicity. The dynamics of such a system is described by Eqs. (1) and (2) in the main text. Here, the matrix $J_{ij}$ is an $N$ by $N$ matrix, where $N$ is the number of DOPO neurons (i.e., the number of DOPO pulses is $2N$).

*Kuramoto model.* Now we consider the following standard Kuramoto model[4,5], which helps us to understand one of the essential points of the synchronization of the DOPO neurons:

$$\frac{d\theta_i}{dt} = \omega_i - J_K \sum_j \sin(\theta_i - \theta_j), \qquad \text{(D1)}$$

where node-dependent $\omega_i$ is assumed to be a natural distribution with a variance $\sigma_\omega$. In this well-known model, the synchronization can be understood from the analogy to the phase transition with an order parameter $re^{i\psi} = \frac{1}{N}\sum_j e^{i\theta_j}$.

We here discuss the analogy between the DOPO neuron model and the Kuramoto model, and then we will discuss about the difference below. Here we set $\gamma = \gamma'(\equiv J_K)$ and $J_{ij} = 1 - \delta_{ij}$. This $J_{ij}$ represents that the DOPO neurons have connections on a complete graph structure. By assuming that

$i$-dependence on $R_i$ is negligibly small, we can rewrite the above equation as

$$\frac{d\theta_i}{dt} = \omega_0 + \frac{R}{4}\sin 4\theta_i - J_K \sum_j \sin(\theta_i - \theta_j), \tag{D2}$$

$$\frac{dR}{dt} = 2P_i R - \frac{R^2}{2}(\cos 4\theta_i + 3) + 2J_K R \sum_{j\neq i} \cos(\theta_i - \theta_j). \tag{D3}$$

As mentioned above, we can roughly evaluate the frequency of an oscillator $\omega_i(P_i) = \omega_0\sqrt{1 - \frac{P_i^2}{8\omega_0^2}}$. Finally, we obtain the effective Kuramoto model given by

$$\frac{d\theta_i}{dt} = \omega_i(P_i) - J_K \sum_j \sin(\theta_i - \theta_j). \tag{D4}$$

As discussed later, we should note that the frequencies $\omega_i(P_i)$ include the effects of the synchronization.

*Pump renormalization*. Next we discuss the pump renormalization effects, which are characteristic of the synchronization of the DOPO neurons. From Eq. (D3) we can infer that the pump amplitude should be renormalized as $\tilde{P}_i' = \tilde{P}_i + J_K \sum_{j\neq i} \cos(\theta_i - \theta_j)$. By using the order parameter of the Kuramoto model defined as $re^{i\psi} = \frac{1}{N}\sum_j e^{i\theta_j}$, we can obtain $\tilde{P}_i' = \tilde{P}_i + r(N-1)J_K$. It suggests that the synchronization causes an effective change in the pump amplitude, and thus, at the network level, the synchronization induces the crossover between AH and SNLC bifurcations. This is the mechanism of the large shift of the spiking frequency after synchronization as discussed in the main text.

To confirm the mechanism above, we performed numerical simulations of coupled 100 DOPO neurons. Figure D1 shows the change in spiking frequencies caused by the synchronization. The left panel shows frequencies as functions of an interaction strength $J_K$ and also an effective pump amplitude $P'$, where the change in frequencies of the single DOPO neuron as a function of the original (not renormalized) pump amplitude $P$ is also shown for comparison. The right panel shows histograms of frequencies for $NJ_K/\omega_0$= 0, 0.1, 0.4,...,1.6.

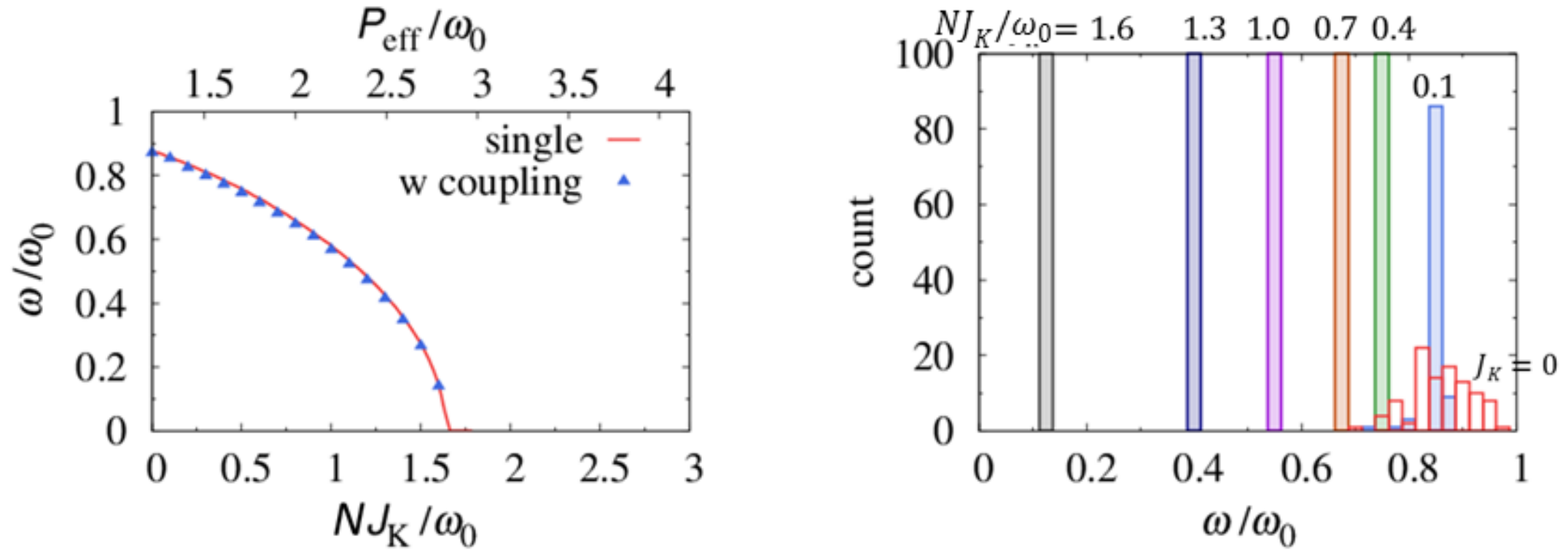

**Fig. D1** Firing rate $\tilde{\omega}$ (on average over the neuronal indices) as a function of $\tilde{J}_K$ or $\tilde{P}$, and histogram of $\tilde{\omega}$ for $NJ_K/\omega_0$ =0, 0.1,0.4,...,1.6. (left) For comparison, we also show spiking frequencies for a single DOPO neuron that were obtained by numerical simulation.

## Section S- E: Additional experimental data

As discussed in the main text, we performed experiments to observe synchronization phenomena of networked DOPO neurons, which demonstrated the pump renormalization effects mentioned in section S-D. Network of 60 DOPO neurons was constructed as depicted in Fig. 3a, in which 15 neurons form an all-to-all connected cluster, and four such clusters were sparsely connected. This network structure was encoded into connections of both the $v$- and $w$-DOPOs (that is, $\gamma = \gamma' \equiv J_k$). We implemented four independent sets of such ensembles consisting of 60 DOPO neurons with different coupling strengths ($\widetilde{J_k}$ = 0, 0.025, 0.05, and 0.075). We set $I_{\mathrm{ext}} = 0$ and used uniform ($i$-independent) coupling of $|J_{vw}| = |J_{wv}|$. The pump amplitude $P_i$ was set to generate the distribution of $\omega_i(P_i)$ and assigned the neurons to four 15-neuron clusters labeled A to D in descending order of firing rate. Figure E1b shows the distribution of measured firing rates of 60 DOPO neurons. Without coupling ($J_k$= 0), the firing rates are widely spread by applied pump $P_i$. As $J_k$ increases, mean and variance of the firing rate are decreased. Figure E1a and E1b shows time evolutions of the phase $\theta_i$ of $i$th DOPO neuron and the order parameter $r$ for each cluster ($N$

= 15) and for all the neurons ($N$ = 60), respectively. With a weak coupling at $\widetilde{J}_k$ = 0.025, coupled DOPO neurons show obvious synchronization in each 15-neuron cluster, but the total order parameter still takes low values because the four synchronized clusters have different firing rates. At $\widetilde{J}_k$ = 0.050, the order parameter averaged over all neurons increases almost periodically, and at $\widetilde{J}_k$ =0.075, the four clusters achieve intermittent synchronization.

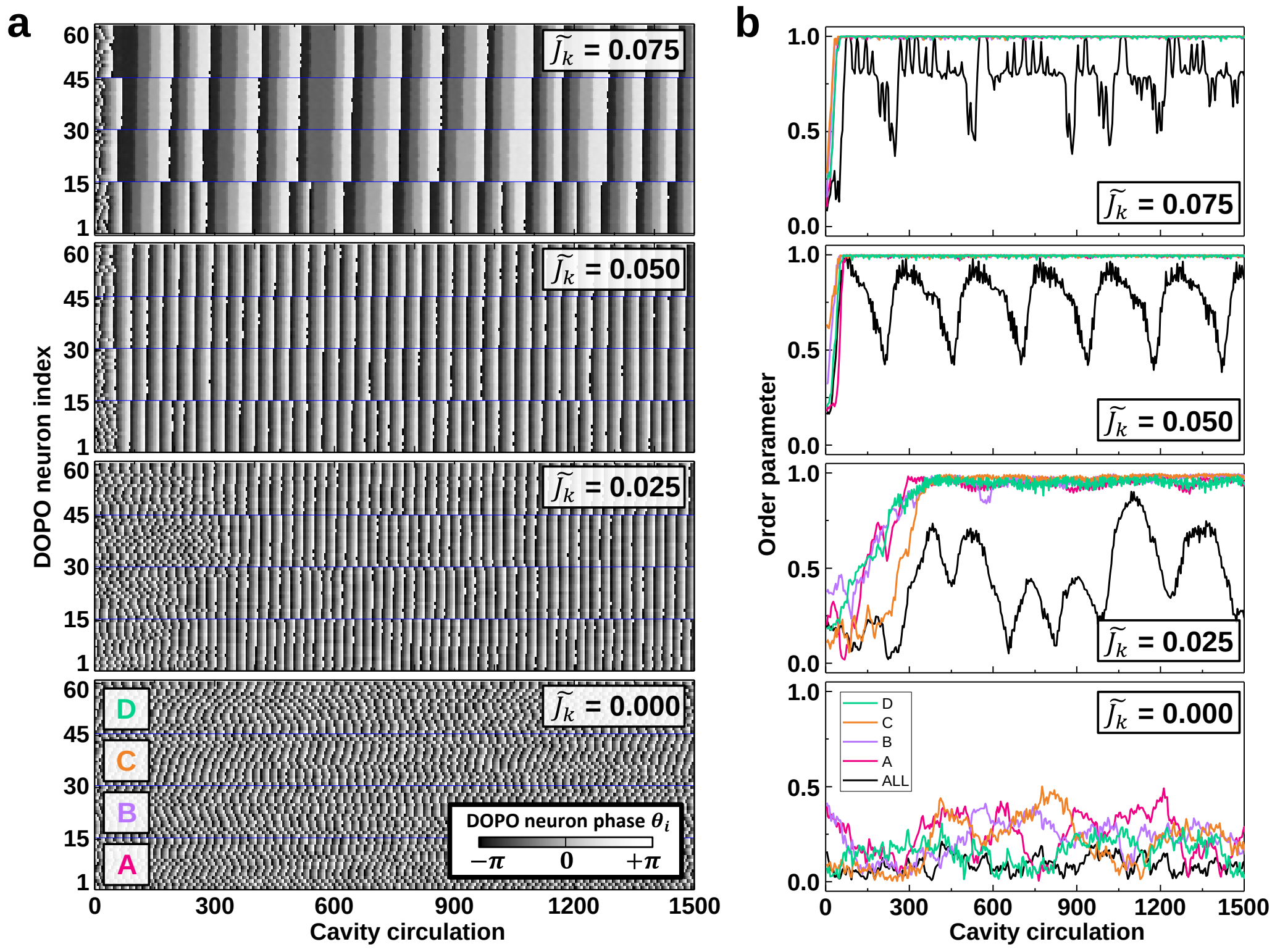

**Fig. E1** Synchronization experiment of clustered Kuramoto models. **a** Time evolutions of phase of the $i$th DOPO neuron. **b** Time evolutions of the order parameter for each cluster and for all 60 neurons.

## Section S-F: Ising model solver

In this section, we discuss the use of the network of DOPO neurons as an Ising model solver. We set $\gamma = -J_K$ and $\gamma' = 0$, and $J_{ij}$ is the spin-spin interaction matrix of the Ising model. Our goal is to minimize the Ising energy. To take account of the mechanism of the control of the firing rates, we define the effective “order” parameter $r_i e^{i\psi_i} = \sum_j \sqrt{\frac{R_j}{R_i}} J_{ij} e^{i\theta_j}$ and introduce $C_i + i\,S_i = r_i e^{i(\psi_i - \theta_i)}$, where $C_i = \sum_j \sqrt{\frac{R_j}{R_i}} J_{ij} \cos(\theta_j - \theta_i)$ and $S_i = \sum_j \sqrt{\frac{R_j}{R_i}} J_{ij} \sin(\theta_j - \theta_i)$. Under the ideal conditions $R_i \sim R_j$ and $\theta_j - \theta_i \sim \{0, \pi\}$, the real part $C_i$ reduces to $\sum_j J_{ij} \cos(\theta_j - \theta_i)$ with $\cos(\theta_j - \theta_i) = \pm 1$, which is equivalent with the local energy $E_{\mathrm{loc},i}$. Note that the present DOPO neuron has the sign-inversion symmetry of $(v, w) \rightarrow (-v, -w)$, and as a result, two symmetric equilibrium points with a phase difference of π usually appear in the $v$-$w$ plane. It suggests that the condition $\theta_j - \theta_i \sim \{0, \pi\}$ is expected to be satisfied in a large $P$ region ($P \geq P_{SN}$), or to be satisfied for a long time in a class-I region (at all times except during the firing).

With $C_i$ and $S_i$, the above equations can be rewritten as

$$\frac{d\theta_i}{dt} = \omega_0' + \frac{R_i}{4} \sin 4\theta_i + \frac{J_K}{2} (S_i \cos 2\theta_i + C_i \sin 2\theta_i), \qquad \text{(F1)}$$

$$\frac{dR_i}{dt} = 2P_i'R_i - \frac{R_i^2}{2}(\cos 4\theta_i + 3) - J_K R_i(C_i \cos 2\theta_i - S_i \sin 2\theta_i), \quad \text{(F2)}$$

where parameters are renormalized as $\omega_0' = \omega_0 - \frac{J_K S_i}{2}$ and $P' = P - \frac{J_K C_i}{2}$. The renormalization of these two parameters yields the change in firing rates. Ideally, energetically stable neurons have large $P'$, suggesting positive correlation of the firing rate and the local energy. In addition, $\cos 2\theta_i$ and $\sin 2\theta_i$ terms change the saddle-node bifurcation points $P_{SNLC}$ (see α dependence of $P_{SNLC}$ in section S-C). The change of the relative value $P/P_{SNLC}$ causes the change of the firing rates because the firing frequency largely changes near $P = P_{SNLC}$. Figure F1 confirms the above discussions in the same way that Fig. D1 does. Note that deviation from the ideal firing rates may become quite large, but the essential qualitative features can be surely captured by the above discussions.

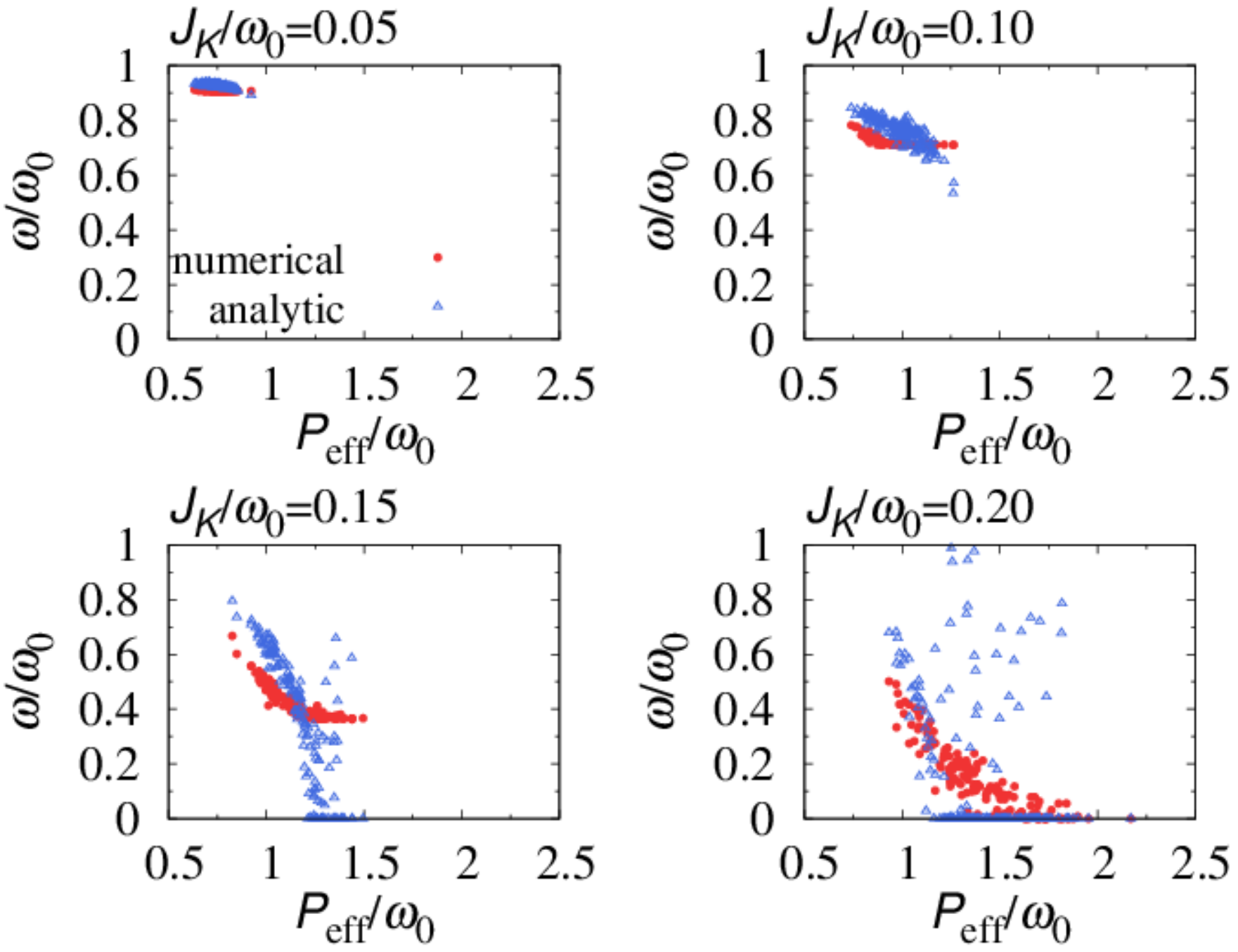

**Fig. F1** Numerically obtained firing rates for the Ising solver with fixed pump amplitude, and firing rates analytically estimated considering renormalization of P and $\omega_0$, which are obtained with the aid of the numerical simulations calculating the average values of $C_i$ and $S_i$.